\documentclass[preprint,showpacs,preprintnumbers,nofootinbib,amsmath,amssymb,aps,prc,longbibliography]{revtex4-1}
\usepackage{graphicx}
\usepackage[colorlinks]{hyperref}
\usepackage{adjustbox}
\usepackage{dcolumn}
\usepackage{bm}
\usepackage{amsmath}
\usepackage{amssymb,amsthm}
\usepackage[intlimits]{mathtools}
\usepackage{multirow}
\usepackage{float}
\usepackage{cases} 
\begin{document}
\title{Renormalizing random-phase approximation by using exact pairing}
\author{L. Tan Phuc$^{1,2}$}
\email{letanphuc191190@gmail.com}
 \author{N. Quang Hung$^{1}$}
 \email{nqhungdtu@gmail.com}
  \author{N. Dinh Dang$^{3}$}
  \email{dang@riken.jp}
 \affiliation{1) Institute of Fundamental and Applied Sciences, Duy Tan University, Ho Chi Minh City
700000, Vietnam\\
2) Faculty of Physics and Engineering Physics, Vietnam National University Ho Chi Minh
City-University of Science, Ho Chi Minh 748355, Vietnam\\
3) Quantum Hadron Physics Laboratory, RIKEN Nishina Center for Accelerator-Based Science, 2-1 Hirosawa, Wako City, 351-0198 Saitama, Japan
}
\date{\today}

\begin{abstract}
A fully self-consistent renormalized random-phase approximation is constructed based on the self-consistent Hartree-Fock mean field plus exact pairing solutions (EP). This approach exactly conserves the particle number and restores the energy-weighted sum rule, which is violated in the conventional renormalized particle-hole random-phase approximation for a given multipolarity. The numerical calculations are carried out for several light, medium, and heavy-mass nuclei such as $ ^{22} $O, $ ^{60} $Ni, and $ ^{90} $Zr by using an effective MSk3 interaction. To study the pygmy dipole resonance (PDR), the calculations are also performed for the two light and neutron-rich $^{24,28}$O isotopes, whose PDRs are known to be dominant. The results obtained show that the inclusion of ground-state correlations beyond the random-phase approximation (RPA) by means of the occupation numbers obtained from the EP affects the RPA solutions within the whole mass range, although this effect decreases with increasing the mass number. At the same time, the anti-pairing effect is observed via a significant reduction of pairing in neutron-rich nuclei. The enhancement of PDR is found in most of neutron-rich nuclei under consideration within our method.
\end{abstract}

\maketitle
\section{Introduction}
\label{Intro}
The random-phase approximation (RPA) is a popular theoretical method to study the low-lying excitations and high-lying giant resonances in nuclei. The RPA states are built on the vibrational collective excitations, which are the superpositions of elementary excitations. The RPA eigenvalues are the energies of the excitations, whereas the isoscalar (IS) and isovector (IV) transition probabilities in a nucleus are calculated by using the components of the RPA eigenvectors.

The RPA excitation operator is composed of many particle-hole (\textit{ph}) components, which are represented by the $ph$ pairs operators: $B_{ph}$ and $B_{ph}^{\dagger}$, where $B_{ph}^{\dagger}=a_{p}^{\dagger}a_{h}$ with $ a_{p}^{\dagger}$ and $ a_{h} $ being the particle ($p$) creation and hole ($h$) annihilation operators, respectively. By assuming that the RPA ground state is not much different from the Hartree-Fock (HF) one and by using the quasi-boson approximation (QBA), the expectation value $ \langle{RPA}\vert{[B_{ph},B_{p'h'}^{\dagger}]}\vert{RPA}\rangle$ of the commutator $[B_{ph},B_{p'h'}^{\dagger}]$ in the RPA ground state is replaced with  that obtained within the HF one, that is $\langle{HF}\vert{[B_{ph},B_{p'h'}^{\dagger}]}\vert{HF}\rangle=\delta_{pp'}\delta_{hh'}$\citep{Ring}. In other words, the QBA implies that the $ph$ pairs behavior like bosons, neglecting their fermionic structure. This is equivalent to the violation of the Pauli principle between the $ph$ pairs. In the region of medium and heavy-mass nuclei, where the nuclear ground-state properties are well described within the HF mean field, the low-lying excitations and giant resonance states are often well described by the RPA. This can be easily understood recalling the fact that the RPA uses the initial inputs from the nuclear mean-field ground state to generate the excitations. However, in light nuclei, the validity of the mean-field description and QBA are still questionable and deserves more study \cite{Ring}. This can be clearly seen especially in exotic light systems, where the existence of clustering within the core is an evidence that the mean-field picture may not hold~\cite{Suzuki}. Also the concept of a nucleon moving in an averaged mean field of the remaining $N-1$ nucleons  is sound only when $N$ is sufficiently large. The mixture of single-particle and collective modes in light nuclei also worsens the QBA.

The restoration of the Pauli principle in the RPA has been carried out within the renormalized RPA (RRPA) by taking into account the ground-state correlations (GSC), which are neglected in the QBA~\cite{Sawicki,Hara,Ikeda,Rowe}. In this method, the expectation value $\langle{RPA}\vert{[B_{ph},B_{p'h'}^{\dagger}]}\vert{RPA}\rangle\simeq D_{ph}\equiv f_h - f_p$ is used instead of the HF one $\langle{HF}\vert{[B_{ph},B_{p'h'}^{\dagger}]}\vert{HF}\rangle=\delta_{pp'}\delta_{hh'}$, which is assumed within the RPA based on the QBA, with $f_k= \langle{RPA}\vert{a_{k}^{\dagger},a_{k}}\vert{RPA}\rangle $ ($k=p,h$)  being the particle ($k=p$) or hole ($k=h$) occupation numbers, respectively, in the correlated RPA ground state $|RPA\rangle$. These occupation numbers $f_k$ can be expressed in term of the RPA eigenvector components, the so-called backward-going amplitude $Y_{ph}^\nu$. The resulting system of the RRPA equations becomes nonlinear with respect to the amplitudes $Y_{ph}^\nu$ in the GSC factor $D_{ph}$, which renormalizes the RPA residual interaction~\cite{Catara,Dukelsky}. These RRPA equations are then self-consistently solved by the iteration.

A major problem of the RRPA, as has been pointed out for the first time in Ref. \cite{Hirsch}, is the violation of various model-independent sum rules, such as the Thomas-Reich-Kuhn (TRK) sum rule for the giant dipole resonance (GDR) or the Ikeda sum rule for the Gamow-Teller transitions, because the GSC factor $D_{ph}$ reduces the absolute values of the matrix elements of the residual interaction. One way to overcome this shortcoming is to take into account the contribution of the particle-particle ($pp$) and hole-hole ($hh$) excitations of all multipolarities~\cite{Hirsch,Catara96,hung2016}. However, this approach is time-consuming as it doubles the size of the RPA matrix. Moreover, although the inconsistency inherent in the QBA is removed by taking into account the particle occupations numbers $f_p>$ 0 and the hole occupation numbers $f_{h}<$ 1, the RRPA still contains another inconsistency as it is still based on the HF mean field, where these occupation numbers are always set to be 0 for all the unoccupied states above the Fermi level, i.e. the $p$ states, and 1 for all the occupied ones below the Fermi level, i.e. the $h$ states.

On the other hand, the RRPA does not include superfluid pairing, which plays an important role, especially in neutron-rich nuclei. Pairing is taken  into account within the quasiparticle RPA (QRPA) \cite{Ring}. However, owing to the QBA for the quasiparticle pair operators (similar to that of the RPA), the standard QRPA also violates the Pauli principle, and the renormalized QRPA (RQRPA) also suffers from the sum rule violation, as has been pointed out in Ref. \cite{Hirsch}. The QRPA also uses the pairing solutions obtained from the Hartree-Fock-Bogoliubov (HFB) or BCS theories, which violate the particle-number conservation, resulting in the chemical potential as a Lagrangian multiplier to be determined in the equation for the average particle number in the ground state. 

In the present paper, we propose a novel approach, which employs the exact pairing solutions (EP)~\cite{Volya2001,Zelevinsky2003}, to renormalize the $ph$RPA. The EP produces the exact occupation numbers $f_k$, which come from the $pp$ and $hh$ pairing correlations. These occupation numbers replace the HF ones, $f_h$ = 1 and $f_{p}=$ 0, in a self-consistent way as has been explained and applied to both self-consistent relativistic \cite{Chen2014} and non-relativistic mean fields \cite{Phuc} in the calculations carried out for realistic nuclei. The Hartree-Fock mean field plus EP (HFEP) supplies a good set of initial inputs for the RRPA, where the GSC factors $D_{ph}$ are obtained in a self-consistent way with the HFEP. In this way, this method resolves three issues at once, namely the above-mentioned inconsistency in the HF mean field used in the conventional RRPA, the inclusion of pairing in the RPA, and the particle number conservation, which is always fulfilled exactly within the EP. The present paper will show if this method of renormalizing the RPA by using the HFEP is capable to restore the Pauli principle and the energy-weighted sum rule (EWSR) without the need of extending the RPA configurations beyond the $ph$ ones. Naturally, the application of this approach requires the monopole pairing correlation in nuclei so that the GSC factors $D_{ph}<$1 can be generated within the EP. The latter, in principle, always generates a finite pairing energy, even for the magic nuclei such as $ ^{48} $Ca \cite{Volya2001}. Therefore, it is expected that our approach can be applied to any nuclei, especially neutron-rich or proton-rich ones.

The proposed approach is applied in calculations for the dipole case in several light, medium, and heavy-mass nuclei, namely $ ^{22} $O, $ ^{60} $Ni, and $ ^{90} $Zr nuclei as well as neutron-rich nuclei $^{24,28} $O, where the pygmy dipole resonance (PDR) has been predicted and/or observed~\cite{Lei,Try,Co1,Co,Hung2013}. 

\section{Formalism}
\subsection{Mean field plus exact pairing}
The ground-state quantities such as the single-particle wave functions $ \varphi $, single-particle energies $ \epsilon $, and nucleon densities $ \rho $, are used as the initial inputs for constructing the excited states in the RPA \cite{Ring}. In this paper, these quantities are extracted from the Skyrme-Hartree-Fock mean field, described by the Hamiltonian
\begin{equation}
\label{Ha0}
\hat{H}_{HF}=\sum_{i}{\hat{t}_{i}}+\sum_{i<j}{v_{ij}}+\sum_{i<j<k}{v_{ijk}} \quad ,
\end{equation}
where $ \hat{t} $ is the kinetic energy, $ v_{ij} $ and $ v_{ijk} $ are the two-body and three-body potentials, respectively. These potentials are included in the expression of the Skyrme interaction as follow
\begin{equation}
\label{pote2}
v_{ij}=t_{0}(1+x_{0}P_{\sigma})\delta{(\vec{r})}+\dfrac{1}{2}t_{1}[\delta(\vec{r})\vec{k}^{2}+\vec{k'}^{2}\delta(\vec{r})]+t_{2}\vec{k'}\delta(\vec{r})\vec{k}+iW_{0}(\vec{\sigma_{i}}+\vec{\sigma_{j}})\vec{k}\times\delta(\vec{r})\vec{k}~,
\end{equation}
\begin{equation}
\label{pote3}
v_{ijk}=t_{3}\delta(\vec{r_{i}}-\vec{r_{j}})\delta(\vec{r_{j}}-\vec{r_{k}})~,
\end{equation}
where $ P_{\sigma}=\dfrac{1}{2}(1+\sigma_{i}\sigma_{j}) $ is the spin-exchange operator expressed via the Pauli spin matrices $\sigma_{i(j)}$, $\vec{k'} $ is the conjugate of the wave vector $ \vec{k} $, and $ \vec{r}=\vec{r_{i}}-\vec{r_{j}} $. The three-body term in the Skyrme interaction can be expressed in terms of the two-body one via the nucleon density \cite{Vautherin1972}
\begin{equation}
\label{pote3-2}
v_{ijk}\longrightarrow v_{ij}=\dfrac{t_{3}}{6}(1+P_{\sigma})\delta(\vec{r_{i}}-\vec{r_{j}})\rho^{\alpha}(\dfrac{\vec{r_{i}}-\vec{r_{j}}}{2}) ~,
\end{equation}
where $\rho=\rho_{Z}+\rho_{N} $ with $\rho_Z$ and $\rho_N$ being the proton and neutron densities, respectively.
\par To include the effect of pairing correlation in the mean field, the Hamiltonian $ \cal H $ of the nuclear system is rewritten in the second quantization \cite{Ring}
\begin{equation}
\label{Ha1}
\hat{\cal H}= \hat{H}_{HF}+\hat{H}_{pair} ~,
\end{equation}
with
\begin{eqnarray}
\label{Hapair}
&& \hat{H}_{HF}= \sum_{j}\epsilon_{j}a_{jm}^{\dagger}a_{jm}~, \\
&& \hat{H}_{pair}=-G\sum_{m m'}a_{jm}^{\dagger}a_{j\widetilde{m}}^{\dagger}a_{j'\widetilde{m}'}a_{j'm'} ~,
\end{eqnarray}
where $a_{jm}^{\dagger} $ and $ a_{jm}$ are the single-particle creation and annihilation operators of a nucleon moving on the $ j $th single-particle levels with projections $ \pm m $ and $G$ is the parameter of the constant monopole pairing interaction. The total (ground-state) energy of the nuclear system is given as
\begin{equation}
\label{Etot}
E= E_{HF}+E_{pair}-E_{c.m.} ~,
\end{equation}
where $ E_{HF} $ and $ E_{pair} $ are the HF and pairing energies, respectively. The correction for the center of mass (c.m.) energy $ E_{c.m.} $, which is presented in detail in Refs. \cite{Langanke,Bender1999}, is subtracted \textit{a posteriori} after the variation of HF equation. The $ E_{pair} $ is obtained by diagonalizing $ \cal H $. The diagonal and off-diagonal matrix elements of this Hamiltonian within the EP are obtained as \cite{Volya2001}
\begin{eqnarray}
&&\left\langle\left\{s_j\right\},\left\{N_j\right\}\left|{\cal H}\right|\left\{s_j\right\},\left\{N_j\right\}\right\rangle = \sum_j\left( \epsilon_j N_j - \frac{G}{4} \left(N_j - s_j \right) \left(2\Omega_j-s_j-N_j + 2 \right) \right)~, \\
&&\left\langle\left\{s_j\right\},...N_j+2,...N_{j'}-2,...\left|{\cal H}\right|\left\{s_j\right\},...N_j,...N_{j'},...\right\rangle \nonumber \\
&&=-\frac{G}{4}\left[\left(N_{j'}-s_{j'}\right)\left(2\Omega_{j'}-s_{j'}-N_{j'}+2\right) \left(2\Omega_{j}-s_{j}-N_{j}\right)\left(N_{j}-s_{j}+2\right)\right]^{1/2}~.
\label{matrix}
\end{eqnarray}
Each basis state $ \vert{\lbrace{s_j}\rbrace},\lbrace{N_j}\rbrace\rangle $ in the matrix elements above represents the $ j $th level with $ N_j=2\Omega_j=2(j+1/2) $ nucleon and $ s_j $ unpaired particles. The pairing energy $ E_{pair} $ and single-particle occupation number $ f_j $, which are obtained after diagonalizing $ \cal H $, are employed to re-define the currents and densities \cite{Phuc,Langanke} and calculate the pairing gap by using the following equations
\begin{align}
\label{quantities1}
& \rho_{q}(r)=\sum_{j}f_{j}\dfrac{2j+1}{4\pi}\varphi_j(r)^2 ~, \\
\label{quantities2}
& \tau_{q}(r)=\sum_{j}f_{j}\dfrac{2j+1}{4\pi}\left[[\partial_{r}\varphi_j(r)]^2+\dfrac{l(l+1)}{r^2}\varphi_j(r)^2\right] ~, \\
\label{quantities3}
& J_{q}(r)=\sum_{j}f_{j}\dfrac{2j+1}{4\pi}[j(j+1)-l(l+1)-\dfrac{3}{4}]\dfrac{2}{r}\varphi_j(r)^2 ~,\\
\label{gap}
& \Delta=\sqrt{-GE_{pair}} ~,
\end{align}
where $ \rho_{q},~\tau_{q},~J_{q} $, and $ \Delta $ are the nucleon densities, kinetic energy densities, spin-current densities, and exact pairing gap, respectively. The subscript $ q $ denotes proton or neutron, and $ \varphi_j $ is the single-particle wave function. Without pairing, the values $ f_j $ in Eqs. (\ref{quantities1}) -- (\ref{quantities3}), which are denoted as $ f^{HF}_j $, are always equal to 1 for the levels below the Fermi surface and 0 for those above it as in the case of HF mean field. With pairing, the values $ f_j $ follow the distribution of exact pairing solutions, namely $ f^{EP}_j<1 $ for the levels below Fermi surface and $ f^{EP}_j>0 $ for those above the Fermi one. These occupation numbers are again used in the currents and densities in Eqs. (\ref{quantities1}) -- (\ref{quantities3}) to re-define them for the initial input of the next step within the RRPA. 
\subsection{Renormalizing random-phase approximation by using exact pairing}
\subsubsection{The phRRPA}
As has been mentioned above, the ground-state quantities are used to construct the RPA excited states. The details of the RPA and RRPA were presented in Ref. \cite{hung2016}. In this section, we will present briefly the main results of these methods. 

The RPA phonon operator is a superposition of $ph$-pair operators in the form \cite{Ring}
\begin{equation}
Q_{JMi}^{\dagger} = \sum_{ph}\bigg[X_{ph}^{Ji}B_{ph}^{\dagger}(JM)- Y_{ph}^{Ji}B_{ph}(J\widetilde{M})\bigg]~,
\label{Q}
\end{equation}
where $ J^{\pi} =0^{+},1^{-},2^{+},... $ is the angular momentum (multipolarity) with natural parity $\pi$, and $M=-J,-J+1,...,J-1,J$ are its projections. The symbol $~\widetilde{}~$ denotes the time-reversal operator ${\cal O}_{J\widetilde M} = (-1)^{J-M}{\cal O}_ {J,-M}$. The operator $B_{ph}^{\dagger}(JM)$ is the $ph$-pair creation operator with the total angular momentum $J$ and projection $M$
\begin{equation}
B_{ph}^{\dagger}(JM) = \sum_{m_pm_h}\langle j_pm_p j_hm_h|JM\rangle a^{\dagger}_{j_pm_p}a_{j_h\widetilde{m}_h}~.
\label{Bph}
\end{equation}
The RPA excited state is defined by acting the phonon operator (\ref{Q}) on the RPA ground stated $ \vert RPA \rangle $, namely
\begin{equation}
|JMi\rangle = Q_{JMi}^{\dagger}|RPA \rangle~,
\label{JMi}
\end{equation}
where $ Q_{JMi}|RPA \rangle = 0$. These RPA states are orthonormalized, viz.
\begin{equation}
\langle JMi|J'M'i'\rangle = \langle RPA|[Q_{JMi},Q_{J'M'i'}^{\dagger}]|RPA\rangle = \delta_{JJ'}\delta_{MM'}\delta_{ii'}~.
\label{ortho}
\end{equation}
The expectation value of the commutation relation $ [B_{ph},B_{p'h'}^{\dagger}]$ in the RPA ground state is calculated as
\begin{eqnarray}
\label{<BB>}
&&\langle RPA|[B_{ph}(JM),B_{p'h'}^{\dagger}(J'M')]|RPA\rangle = \nonumber \\
&&\delta_{j_pj_p'}\sum_{m_pm_hm_h'}\langle j_pm_pj_hm_h|JM\rangle
\langle j_pm_pj_h'm_h'|J'M'\rangle\langle RPA| a^{\dagger}_{j_h\widetilde{m}_h}a_{j_h'\widetilde{m}_h'}|RPA\rangle \nonumber \\
&&- \delta_{j_hj_h'}\sum_{m_pm_p'm_h}\langle j_pm_pj_hm_h|JM\rangle
\langle j_p'm_p'j_hm_h|J'M'\rangle \langle RPA|a^{\dagger}_{j_p'{m}_p'}a_{j_p{m}_p}|RPA\rangle \nonumber \\
&&\simeq\delta_{JJ'}\delta_{MM'}\delta_{j_pj_p'}\delta_{j_hj_h'}D_{ph}~,
\end{eqnarray}
where  $D_{ph}$ is the the GSC factor
\begin{equation}
 D_{ph} \cong f_h - f_p = \langle RPA|a^{\dagger}_{j_h{m}_h}a_{j_h{m}_h}|RPA\rangle - \langle RPA|a^{\dagger}_{j_p{m}_p}a_{j_p{m}_p}|RPA\rangle~.
 \label{GSCDph}
 \end{equation}
In order to obtain a set of linear equations with respect to the amplitudes $X_{ph}^{Ji}$ and $Y_{ph}^{Ji}$ in Eq. (\ref{Q}), it is assumed that the RPA ground state
$ \vert RPA \rangle $ is not much different from the HF one $ \vert HF \rangle $ so that it can be replaced with the latter \cite{Ring}. This leads to 
the substitution $f_h\simeq f^{HF}_h=\langle HF|a^{\dagger}_{j_h{m}_h}a_{j_h{m}_h}|HF\rangle=1 $ and $f_p\simeq f^{HF}_p=\langle HF|a^{\dagger}_{j_p{m}_p}a_{j_p{m}_p}|HF\rangle=0 $. Consequently, the GSC factor $D_{ph}$ is replaced with $ D^{HF}_{ph}=f^{HF}_h - f^{HF}_p=1 $ and the expectation value of the commutation relation (\ref{<BB>}) becomes
\begin{equation}
\langle HF|[B_{ph}(JM),B_{p'h'}^{\dagger}(J'M')]|HF\rangle =\delta_{JJ'}\delta_{MM'}\delta_{j_pj_p'}\delta_{j_hj_h'}~.
\label{QBA}
\end{equation}
This approximation is called the QBA as it treats the $ ph $ pairs like bosons, which obey the exact boson commutation relation. This also means that the QBA ignores (or violates) the Pauli principle between the fermion pairs. Within the QBA, the orthonormal condition (\ref{ortho}) requires the amplitudes $X_{ph}^{Ji}$ and $Y_{ph}^{Ji}$ to obey the following normalization condition
\begin{equation}
\sum_{ph}(X_{ph}^{Ji}X_{ph}^{J'i'} - Y_{ph}^{Ji}Y_{ph}^{J'i'}) = \delta_{JJ'}\delta_{ii'}~,
\label{norm}
\end{equation} 
whereas the closure relations
\[
\sum_{i}(X_{ph}^{Ji}X_{p'h'}^{Ji} - Y_{ph}^{Ji}Y_{p'h'}^{Ji}) = \delta_{pp'}\delta_{hh'}~, 
\]
\begin{equation}
\sum_{i}(X_{ph}^{Ji}Y_{p'h'}^{Ji} - Y_{ph}^{Ji}X_{p'h'}^{Ji}) = 0~, 
\label{closure}
\end{equation}
ensure the inverse expression of $ph$-pair creation operator in terms of the phonon one
\begin{equation}
B_{ph}^{\dagger}(JM) = \sum_{i}[X_{ph}^{Ji}Q_{JMi}^{\dagger} + Y_{ph}^{Ji}Q_{J\widetilde{M}i}]~.
\label{B}
\end{equation}

By using the boson-mapping technique~\cite{Catara,Rowe,hung2016} to express the particle-number operator in terms of the sums of products $B_{ph}^{\dagger}B_{ph}$, the particle and hole occupation numbers are calculated within the RPA as
\begin{eqnarray}
&& f_p^{RPA} = \frac{1}{2j_p+1}\langle HF|\sum_{m_p}a^{\dagger}_{j_pm_p}a_{j_pm_p}|HF\rangle=\frac{1}{2j_p+1}\sum_{Ji}(2J+1)\sum_h(Y_{ph}^{Ji})^2~, \\
&& f_h^{RPA} = 1-\frac{1}{2j_h+1}\langle HF|\sum_{m_h}a_{j_hm_h}a^{\dagger}_{j_hm_h}|HF\rangle = 1 - \frac{1}{2j_h+1}\sum_{Ji}(2J+1)\sum_p(Y_{ph}^{Ji})^2~.
\label{fpfh}
\end{eqnarray}

The RRPA phonon operators are different from the RPA ones by the presence of the GSC factor $D_{ph}$, which is smaller than 1~\cite{Catara}
\begin{equation}
{\cal Q}_{JMi}^{\dagger} = \sum_{ph}\bigg[\frac{{\cal X}_{ph}^{Ji}}{\sqrt{D_{ph}}}{B}_{ph}^{\dagger}(JM)- \frac{{\cal Y}_{ph}^{Ji}}{\sqrt{D_{ph}}}{B}_{ph}(J\widetilde{M})\bigg]~,
\label{RQ}
\end{equation}
and
\begin{equation}
B_{ph}^{\dagger}(JM) = \sqrt{D_{ph}}\sum_{i}[{\cal X}_{ph}^{Ji}{\cal Q}_{JMi}^{\dagger} + {\cal Y}_{ph}^{Ji}{\cal Q}_{J\widetilde{M}i}]~.
\label{RB}
\end{equation}
In this form, the RRPA amplitudes ${\cal X}_{ph}^{Ji}$ and ${\cal Y}_{ph}^{Ji}$ fulfill the same normalization and closure relations as those of RPA [Eqs. (\ref{norm})--(\ref{closure})]. However, the occupation numbers within the RRPA are now calculated from the recurrent expressions~\cite{Catara,hung2016}
\begin{equation}
f_p = \frac{1}{2j_p+1}\langle RPA|\sum_{m_p}a^{\dagger}_{j_pm_p}a_{j_pm_p}|RPA\rangle= \frac{1}{2j_p+1}\sum_{Ji}(2J+1)\sum_hD_{ph}({\cal Y}_{ph}^{Ji})^2~, 
\label{fp}
\end{equation}
\begin{equation}
f_h = 1-\frac{1}{2j_h+1}\langle RPA|\sum_{m_h}a_{j_hm_h}a^{\dagger}_{j_hm_h}|RPA\rangle = 1 - \frac{1}{2j_h+1}\sum_{Ji}(2J+1)\sum_pD_{ph}({\cal Y}_{ph}^{Ji})^2~,
\label{fh}
\end{equation}
with 
\begin{equation}
D_{ph} \equiv f_h - f_p =1 - \sum_{Ji}(2J+1)\bigg[\frac{1}{2j_p+1}\sum_{h'}D_{ph'}({\cal Y}_{ph'}^{Ji})^2+\frac{1}{2j_h+1}\sum_{p'}D_{p'h}({\cal Y}_{p'h}^{Ji})^2\bigg]~.
\label{Dph}
\end{equation}
The amplitudes ${\cal X}_{ph}^{Ji}$ and ${\cal Y}_{ph}^{Ji}$ are calculated based on the components of the eigenvectors 
of the $ph$RRPA matrix equation
\begin{equation}
\left( \begin{array}{cc}
A & B  \\
-B & -A \end{array} 
\right)\left( \begin{array}{c}
{\cal X}^{Ji} \\
{\cal Y}^{Ji}\end{array} \right)
={E}_{Ji}\left( \begin{array}{c}
{\cal X}^{Ji} \\
{\cal Y}^{Ji}\end{array} \right)~,
\label{RRPA}
\end{equation}
where $ {E}_{Ji} $ are the $ph$RRPA eigenvalues (phonon energies). The matrices A and B are given as
\begin{eqnarray}
\label{A}
&& A_{ph, p'h'} = (\epsilon_p - \epsilon_h)\delta_{pp'}\delta_{hh'} +\sqrt{D_{ph}D_{p'h'}}\langle ph'|V_{res}|hp'\rangle~,\hspace{5mm} \\
&& B_{ph, p'h'} = \sqrt{D_{ph}D_{p'h'}}\langle pp'|V_{res}|hh'\rangle~,
\label{B}
\end{eqnarray}
where $ \epsilon_k $ is the single-particle energy of a spherical orbital $|j_k,m_k\rangle$ with $k = p, h$ and $ V_{res} $ is the two-body residual interaction \cite{Colo2013}. The presence of the GSC factors $ \sqrt{D_{ph}D_{p'h'}}$ renormalizes the residual interaction by reducing the absolute value of its matrix element $\langle ph'|V_{res}|hp'\rangle$ to $\sqrt{D_{ph}D_{p'h'}}\langle ph'|V_{res}|hp'\rangle$. The $ph$RRPA equations are nonlinear with respect to the amplitudes ${\cal Y}_{ph}^{Ji}$ and need to be solved by iteration. In the first step, the RPA equations with $D_{ph}=1$ are solved. The GSC factor $D_{ph}$ is then calculated by using the RPA occupation numbers $f_h$ and $f_p$ defined in Eq. (\ref{fpfh}). The $ph$RRPA matrix equation (\ref{RRPA}) is then diagonalized to obtain a new set of eigenvectors, which produces new GSC factors $D_{ph} $ for the next step. This process is repeated self-consistently until the criterion of convergency is achieved.
\subsubsection{Inclusion of exact pairing}
As has been mentioned previously, the collectivity and EWSR are reduced within the $ph$RRPA \cite{Catara,hung2016}. One way to remove this drawback is extending the $ph$RRPA to include the $ pp $ and $ hh $ configurations on the same footing with the $ph$ ones for all multipolarities \cite{Catara,Catara96,Hirsch,hung2016}. However, this leads to a significant expansion of the RPA matrix and solving the RRPA equations becomes time-consuming. In the present paper, we propose an alternative method by using the exact solutions of the pairing problem (Sec. II. A) to renormalize the $ ph $RPA. We expect that not only this method can restore the EWSR at each multipolarity $J^{\pi}$ without the need of including the $ pp $ and $ hh $ excitations, but it also takes into account the exact pairing, having ensured the exact particle number already in the reference state, unlike the HFB or BCS theories used in (R)QRPA. The Hamiltonian of the nuclear system is now written as
\begin{equation}
\label{Hatt}
\hat{\cal H}= \hat{H}_{HF}+\hat{H}_{pair}+\hat{H}_{res} ~,
\end{equation}
where $ \hat{H}_{res} $ is the residual Hamiltonian~\cite{Ring}:
\begin{equation}
\label{Hares}
\hat{H}_{res}= \sum_{php'h'}A_{php'h'}B_{ph}^{\dagger}B_{p'h'}+\dfrac{1}{2}\sum_{php'h'}(B_{php'h'}B_{ph}^{\dagger}B_{p'h'}+h.c)~.
\end{equation}
Because the size of the matrix to be diagnolized in the exact pairing Hamiltonian is limited~\cite{Hung2009,Hung2017}, only the levels in a truncated spectrum around the Fermi surface is used for the EP. The occupation number of the hole and particle states outside the truncated space ${\cal T}$ remain to be 1 and 0, respectively. These exact occupation numbers are used to produce the GSC factors (\ref{DphEPa}) and (\ref{DphEPb}) which are employed to renormalize the RPA as mentioned in Sec. II.B.1. In particular, all the GSC factor $D_{ph}$ in the $ph$RRPA matrix (\ref{A}) and (\ref{B}) are now replaced with ${D}_{ph}^{EP}$ as 
\begin{subnumcases}{D_{ph}^{EP}=}
\label{DphEPa} 
f_h^{EP}-f_p^{EP} & ($p,h \in \lbrace{\cal T}\rbrace$)~, \\
1 & ($p,h \notin \lbrace{\cal T}\rbrace$)~,
\label{DphEPb} 
\end{subnumcases}
and the set of $ph$RRPA equations is diagonalized. The backward-going amplitudes ${\cal Y}_{ph}^{Ji}$, obtained after this diagonalization, are used to calculate the RPA occupation numbers $ f_k^{RPA} $ ($ k=p, h $), following Eqs. (\ref{fp}) and (\ref{fh}) for each multipolarity $ J^{\pi} $ with $ \pi=(-1)^J $. These occupation numbers are then used to replace those within the Hartree-Fock mean field as the initial values of the new loop. These steps are repeated until the convergence is reached, namely each single-particle energy satisfies the criterion $ \vert \epsilon_j(n)-\epsilon_j(n-1)\vert \leqslant 10^{-4} $ MeV. 

The total energy of the nuclear system is calculated as
\begin{equation}
\label{Ett}
E=E_{HF}+E_{pair}+E_{RPA}-E_{c.m.}
\end{equation}
where the RPA energy $ E_{RPA} $ is given as \cite{Ring}
\begin{equation}
\label{ERPA}
E_{RPA}=-\dfrac{1}{2}{\rm Tr}A+\dfrac{\hbar}{2}\sum_{Ji}E_{Ji}=-\sum_{Ji}\hbar E_{Ji}\sum_{ph}\dfrac{\vert{\cal Y}_{ph}^{Ji}\vert^2}{D_{ph}}.
\end{equation}
The pairing-strength parameter $ G $ in the EP calculation is adjusted so that the EWSR is fulfilled and the pairing gap obtained within the exact pairing is close to the experimental odd-even mass difference. This procedure guarantees a full consistency between the mean field and the renormalization process using EP. We refer to this method as the SC-HFEPRPA hereafter.
\subsubsection{The EWSR}
The reduced transition probabilities $B(EJ)$ between the ground state $|0\rangle$ and excited state $|\nu \rangle\equiv |JMi \rangle $ within the SC-HFEPRPA have the same form as that in the conventional $ph$RRPA, namely
\begin{equation}
B(EJ, 0\rightarrow \nu)(E_{Ji}) = | \langle\nu |\hat{F}_{J}|0\rangle|^2=\bigg|\sum_{ph}\sqrt{D_{ph}}({\cal X}_{ph}^{Ji} + {\cal Y}_{ph}^{Ji})\langle p ||\hat{F}_J||h\rangle\bigg|^2~,
\label{BE}
\end{equation}
where $\langle p||\hat{F}_J||h\rangle$ are the reduced matrix elements of the one-body excitation operators $ \hat{F}_{JM} $ \cite{Bohr, Colo2013}. To present the distribution of the probabilities $ B(EJ, 0\rightarrow \nu)(E_{Ji}) $ over the discrete one-phonon states $|\nu\rangle$ with energies $E_{Ji}$ as a continuous function of the excitation energy $E$, this distribution is often smoothed by representing the delta function as $\delta(x) = \varepsilon/[\pi(x^2+\varepsilon^2)]$. As the result, one obtains the strength function
\begin{equation}
S_{J}(E) = \frac{\varepsilon}{\pi}\sum_i\frac{B(EJ, E_{Ji})}{(E-E_{Ji})^2+\varepsilon^2}~,
\label{strength}
\end{equation}
where $ \varepsilon $ is the smoothing parameter. This parameter sometime is associated with the escape width $\Gamma^{\uparrow}\equiv2\varepsilon$ caused by the coupling to the continuum, which is around few hundred keV \cite{hung2016}, or even with the spreading width $\Gamma^{\downarrow}$ (several MeVs) of the giant resonance, which is caused by coupling of $1p1h$ to more complicate configurations such as $2p2h$ etc, as the mechanisms are beyond the reach of the RPA. The energy-weighted sum of strength (EWSS) of the $B(EJ)(E_{Ji})$ distribution at each multipolarity $ J^{\pi}$ is obtained as the sum of $B(EJ)$ with the weight $E_{Ji}$ or the integral of $E\times S_{J}(E)$ within the energy interval $0\leq E (E_{Ji}) \leq E_{max}$. As the electric dipole excitations ($J^{\pi} = 1^{-}$) are considered in this paper, the $E1$ EWSS within the SC-HFEPRPA is given as the first moment
\begin{equation}
m_1 = \int_0^{E_{max}}ES_1(E)dE = \sum_{i}E_{i}B(E1, E_{i})~,
\label{m1}
\end{equation}
where the multipolarity $J=1$ in the subscript for $E_{Ji}$ is omitted for simplicity.

\begin{figure}[h]
       \includegraphics[scale=0.5]{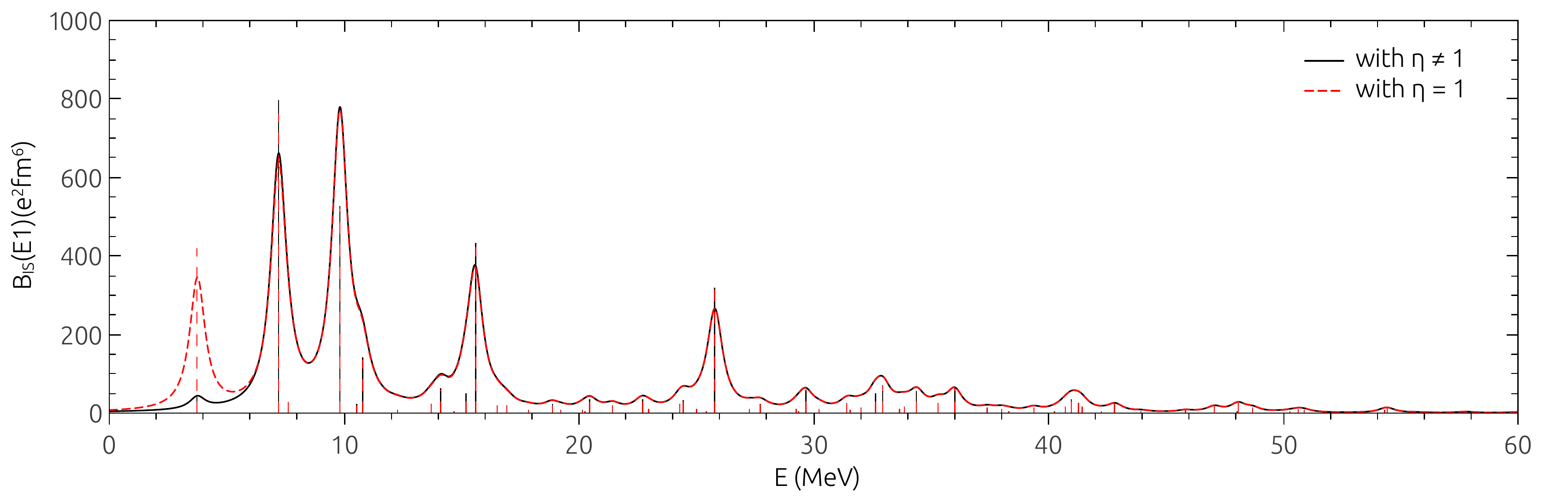}
       \caption{(Color online) The IS dipole transition probabilities of $ ^{22} $O obtained within the SC-HFEPRPA with and without $\eta$. 
        \label{eta}}
    \end{figure}

For a complete set of the exact eigenstates $|\nu\rangle$, the EWSR holds, for which the first $m_1 $ (\ref{m1}) is equal to half the expectation value of the double commutator $[\hat{F},[H,\hat{F}]]$ in the ground state, that is $m_1=\dfrac{1}{2}\langle 0|[\hat{F},[H,\hat{F}]]|0\rangle$.  The standard RPA fulfills the EWSR, where $m_1$ is calculated within the standard RPA, whereas the ground state $|0\rangle$ is replaced with the HF ground state $|HF\rangle$~\cite{Thouless}. In the present paper, the fulfillment of the IS and IV EWSRs is verified based on the ratio of the $m_1$ value calculated within the HF-EP-RPA to its corresponding theoretical value. For the $E1$ resonances, to avoid the overlap between the spurious state and the physical states, the ground-state expectation value of the double commutator is estimated after subtracting the effect of center of mass motion~\cite{Harakeh,Giai,Colo2013}. The $E1$ IS EWSR is obtained in this way as 
\begin{equation}
m_1^{IS} = \frac{\hbar^2}{2m^{*}}\frac{A}{4\pi}(33\langle r^4\rangle - 25\langle r^2\rangle^2) ~,
\label{ISEWSR}
\end{equation}
where $ A $, $ m^*$, and $ \langle r^2\rangle $ are the atomic mass, corected (effective) mass, and root-mean-square radius of the nucleus, respectively. Within the standard HF-RPA, $m^*$ is equal to $mA/(A-1)$ since $f_h^{HF}=1$ and $f_p^{HF}=0$. Within the SC-HFEPRPA, $m^*$ is approximated by using the one-body part of the momentum operator \cite{Langanke}
\begin{equation}
P^2_{c.m.}\approx \sum_jf_j\hat{p}^2_j\approx \overline{f}\sum_j\hat{p}^2_j~,
\label{Pcm}
\end{equation}
where the average occupation number of the hole states is $ \overline{f}=\dfrac{\sum_{j=h}f_j}{\sum_{j=h}j} $. The two-body part of the momentum operator and the average occupation number of the particle states, whose contribution is negligible, are omitted for simplicity, so that the center-of-mass correction can be expressed in a compact form in terms of the nucleon mass as
\begin{equation}
m^*=m\dfrac{A}{A-\overline{f}}~.
\label{m*}
\end{equation}
After the first loop of the iteration, the mean field is modified by the single-particle occupation numbers, that is, $ f^{HF}_j$ is replaced with $f_j^{RPA} $. Because of the center-of-mass motion, which is modified by the EP and RPA occupation numbers, the ground-state expectation values for the nuclear density distribution must be corrected. The latter are usually expressed in terms of their radial moments \cite{Bohr,Alkhazov,Neugart}
\begin{equation}
\langle r^{2n}\rangle =\int \rho(r)r^{2n}d^3r~.
\label{<r>}
\end{equation}
After diagonalizing the EP matrix to obtain the new HF density with EP $ \rho_j^{HFEP}(r) $, the nuclear radial moments are corrected \textit{a posteriori} to contribute to the IS EWSR 
\begin{equation}
\langle r^{2n}\rangle =\int \eta \rho^*(r)r^{2n}d^3r~, \hspace{5mm} \rho^*(r) =\sum_j f_j^{mf}\rho_j^{mf}(r)~,
\label{<r>cor}
\end{equation}
where $ f_j^{mf} $ is the occupation number in mean field, which is equivalent to $f_j^{HF}=1 $ within the self-consistent HF approximation, $f_j^{EP} $ within the self-consistent HFEP, or $f_j^{RPA}$ within the SC-HFEPRPA. The radial moments are renormalized by using the parameter $\eta$  so that Eq. (\ref{<r>}) holds for the first-order radial moment ($ n=0 $) after including pairing and correlations from the residual interactions into the mean field, that is
\begin{equation}
\int_0^\infty \eta \rho^*(r)d^3r=1~.
\label{r0}
\end{equation}
The integrand (\ref{r0}) represents the distribution of one nucleon in the radial mesh of nucleus. This normalization is important as it helps to reduce the transition probability at the spurious state, keeping the total IS EWSS almost unchanged as shown in Fig. \ref{eta}, where the value of the $E1$ IS EWSS obtained for $^{22}$O within the SC-HFEPRPA by using $ \eta\neq$1 (0.677391$\times 10^{5} $ e$^2$fm$^6 $MeV) is larger than that given with $\eta=1$ (0.671324$\times 10^{5} $ e$^2$fm$^6 $MeV) only by about 0.9$\%$.\\
The $E1$ IV EWSR  is expressed in term of the model independent Thomas-Reiche-Kuhn (TRK) sum rule as 
\begin{equation}
m_1^{IV} = (1+\kappa)TRK~, \hspace{5mm}TRK = \frac{9}{4\pi}\frac{\hbar^2}{2m^{*}} \frac{NZ}{A}~.
\label{IVEWSR}
\end{equation}
The enhancement factor $ \kappa $ in Eq. (\ref{IVEWSR}) is caused by the velocity-dependent terms of the Skyrme interactions \cite{Liu}, whose value is given in Eq. (29) of Ref. \cite{Colo2013}.
\section{Results and discussion}
The numerical calculations in the present paper were carried out making use of the self-consistent HF-RPA code with full residual interaction. This code was developed  by Col\`{o} {\it et al.} and made accessible for the nuclear physics community~\cite{Colo2013}. In the present paper, we have extended this code to include the EP and renormalize the RPA residual interactions. The main limitation of this code is that it can be used to calculate the properties of spherical nuclei with the filled sub-shells. This limitation remains within our extension. Based on the test by using a series of BSk and MSk interactions conducted in Ref. \cite{Phuc}, the Skyrme interaction MSk3 is employed in the present paper. The self-consistent HFEP calculations using this MSk3 interaction reproduces well the experimental ground-state properties (binding energy, radii, and nucleon density) of all nuclei under consideration in the present paper. For example, the difference between the calculated average binding energy $BE/A$ and the experimental data is lower than 0.5$\%$. The renormalization of the RPA is proceeded in two ways: 1) The HFEP is solved self-consistently as in Ref. \cite{Phuc}. Then, the RPA equations are solved once in the end. This process is referred to as the non-self-consistent HF-EP-RPA (HFEPRPA) hereafter; 2) The HF, EP, and RPA are iteratively solved for each loop until the convergency is reached. This process is referred to as the SC-HFEPRPA. We consider three spherical nuclei, whose masses range from light to heavy, namely $ ^{22} $O, $ ^{60} $Ni, and $ ^{90} $Zr. The calculations are performed in a box of 15 fm radius with the radial mesh of 0.1 fm and cut-off energy $ E_c =60$ MeV for $ J^\pi=1^{-} $. This choice is reasonable for the calculations for three selected nuclei above \cite{Colo2013}. 

\begin{figure}[h]
       \includegraphics[scale=0.5]{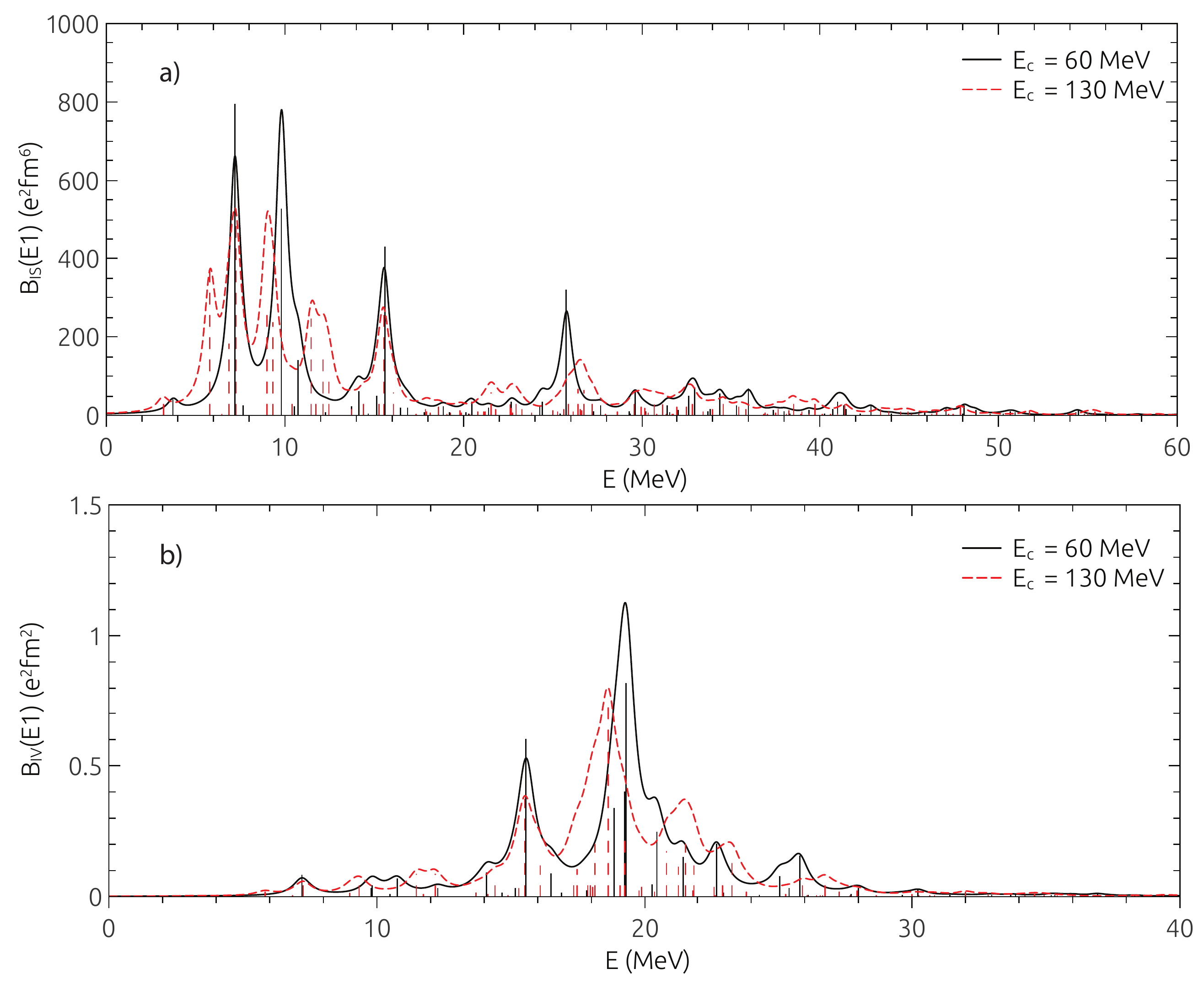}
       \caption{(Color online) The IS (a) and IV (b) dipole transition probabilities of $ ^{22} $O obtained within the SC-HFEPRPA by using different cut-off energies $E_c$. 
        \label{Ec}}
    \end{figure}

\begin{figure}[h]
       \includegraphics[scale=0.5]{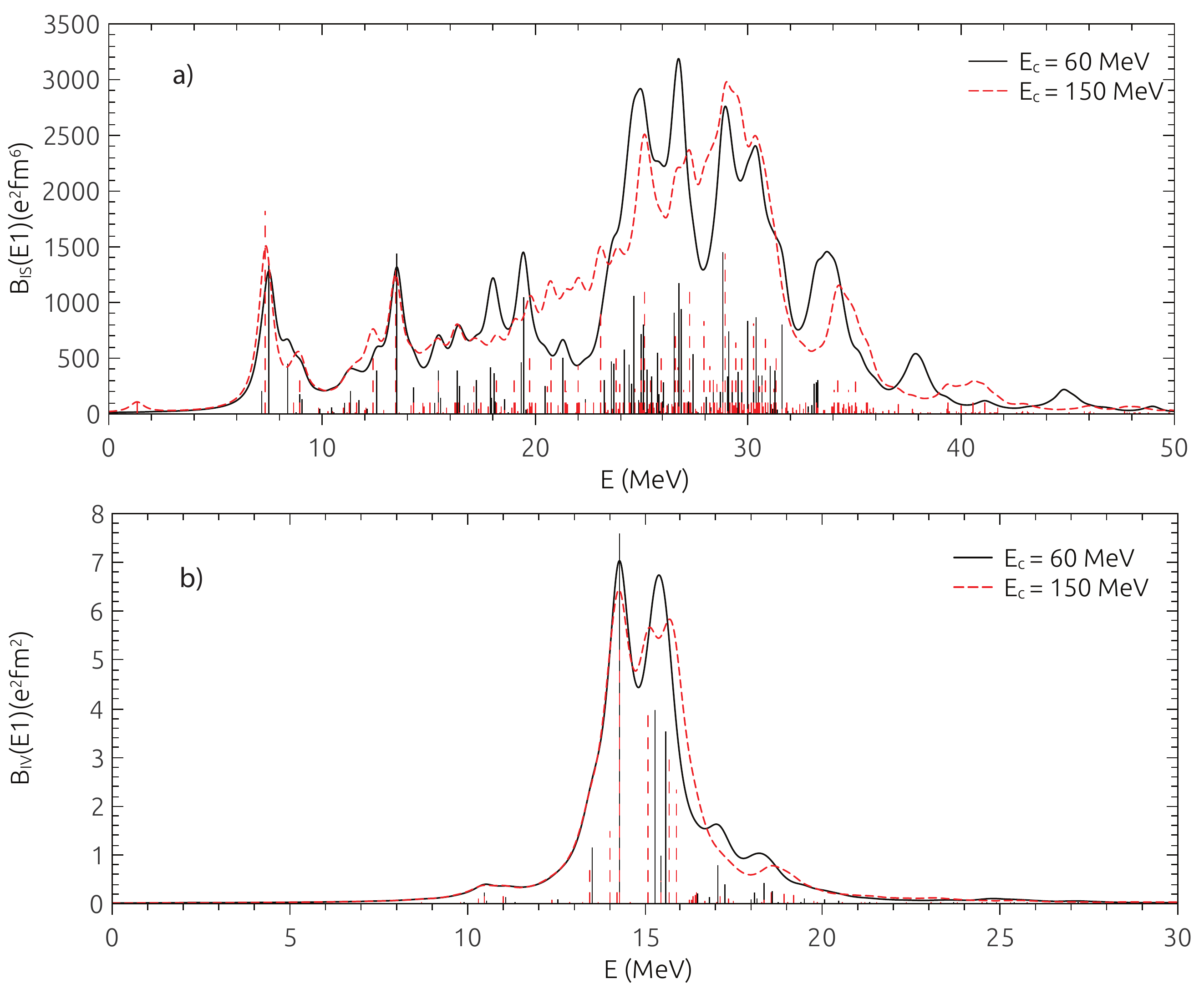}
       \caption{(Color online) Same as Fig. \ref{Ec} but for $ ^{90} $Zr. 
        \label{Ec90}}
    \end{figure}

The multipolarity $ J^\pi=1^{-} $ is used in our calculation, which is the most important for numerically evaluating the E1 EWSR to be compared with the model independent sum rule, that is the TRK one. The cut-off energy $E_c =60$ MeV is sufficiently large to perform our calculation because the spurious state obtained with this $E_c$ is well separated from the physical ones as has been shown in Ref. \cite{Colo2013}. By increasing the cut-off energy $ E_c $ to 130 MeV, we found that the energy of the spurious state in $ ^{22} $O is indeed shifted down from 3.75 MeV to 3.12 MeV (see Fig. \ref{Ec}). The same trend is seen in the results of calculations with $ E_c =60$ and 150 MeV for the heavy $^{90}$Zr nucleus as shown in Fig. \ref{Ec90}. The energy of spurious state in this nucleus reduces from 2.24 MeV to 1.34 MeV. These results show that the higher $E_c$ we choose, the lower spurious energy we can obtain. However, the calculations also become much more time consuming. Therefore, the value $ E_c =60$ MeV is chosen. 


As for the EP calculation, because of the limitation of the size of the pairing matrix to be diagonalized, we cannot carry out the EP calculations within a too large space of single-particle levels. We therefore adopted the truncated space given in Table \ref{table1} for each nucleus. This truncated space is chosen based on an assumption that pairing affects only few single-particle levels around the Fermi surface \cite{Hung2009,Hung2017}. The numbers of nucleons, which are left outside these truncated spaces, are 0($N$)-8($Z$), 8($N$)-28($Z$), and 50($N$)-28($Z$) for $^{22}$O, $^{60}$Ni, and $^{90}$Zr, respectively. They fill the shells, which form the magic cores, hence preventing any correlation. Two additional oxygen isotopes, $ ^{24} $O and $ ^{28} $O, are also employed in the study of the PDR in neutron-rich nuclei by using our method. The GSC factor $ D_{ph} $ is obtained after diagonalizing the EP matrix. The values of RPA occupation number $f_k $ $(k=p,h)$ are obtained by solving Eqs. (\ref{fpfh}) and (\ref{Dph}) self-consistently with the accuracy $\vert D_{ph}(n)-D_{ph}(n-1)\vert \leqslant 10^{-3}$ with $n$ being the number of iterations.

\begin{table}[h]
\caption{The truncated spaces used in the EP calculations for $^{22}$O, $^{60}$Ni, and $^{90}$Zr.}
\begin{tabular}{|c|c|c|c|c|c|c|c|c|c|c|c|c|}
\hline\hline
	&Hole levels&Particle levels\\
\hline
    $^{22}$O (Neutron)& 1s$_{1/2}$, 1p$_{3/2}$, 1p$_{1/2}$, 1d$_{5/2}$ & 2s$_{1/2}$, 1d$_{3/2}$, 2p$_{3/2}$, 2p$_{1/2}$, 3s$_{1/2}$, 2d$_{5/2}$\\
    $^{60}$Ni (Neutron)& 1d$_{5/2}$, 2s$_{1/2}$, 1d$_{3/2}$, 1f$_{7/2}$, 2p$_{3/2}$ & 2p$_{1/2}$, 1f$_{5/2}$, 1g$_{9/2}$ \\
   $^{90}$Zr (Proton)& 2p$_{3/2}$, 1f$_{5/2}$, 2p$_{1/2}$ & 1g$_{9/2}$, 2d$_{5/2}$, 3s$_{1/2}$, 2d$_{3/2}$, 1g$_{7/2}$, 1h$_{11/2}$, 2f$_{7/2}$\\
    \hline\hline
\end{tabular}
\label{table1}
\end{table}

\begin{figure}[h]
       \includegraphics[scale=0.45]{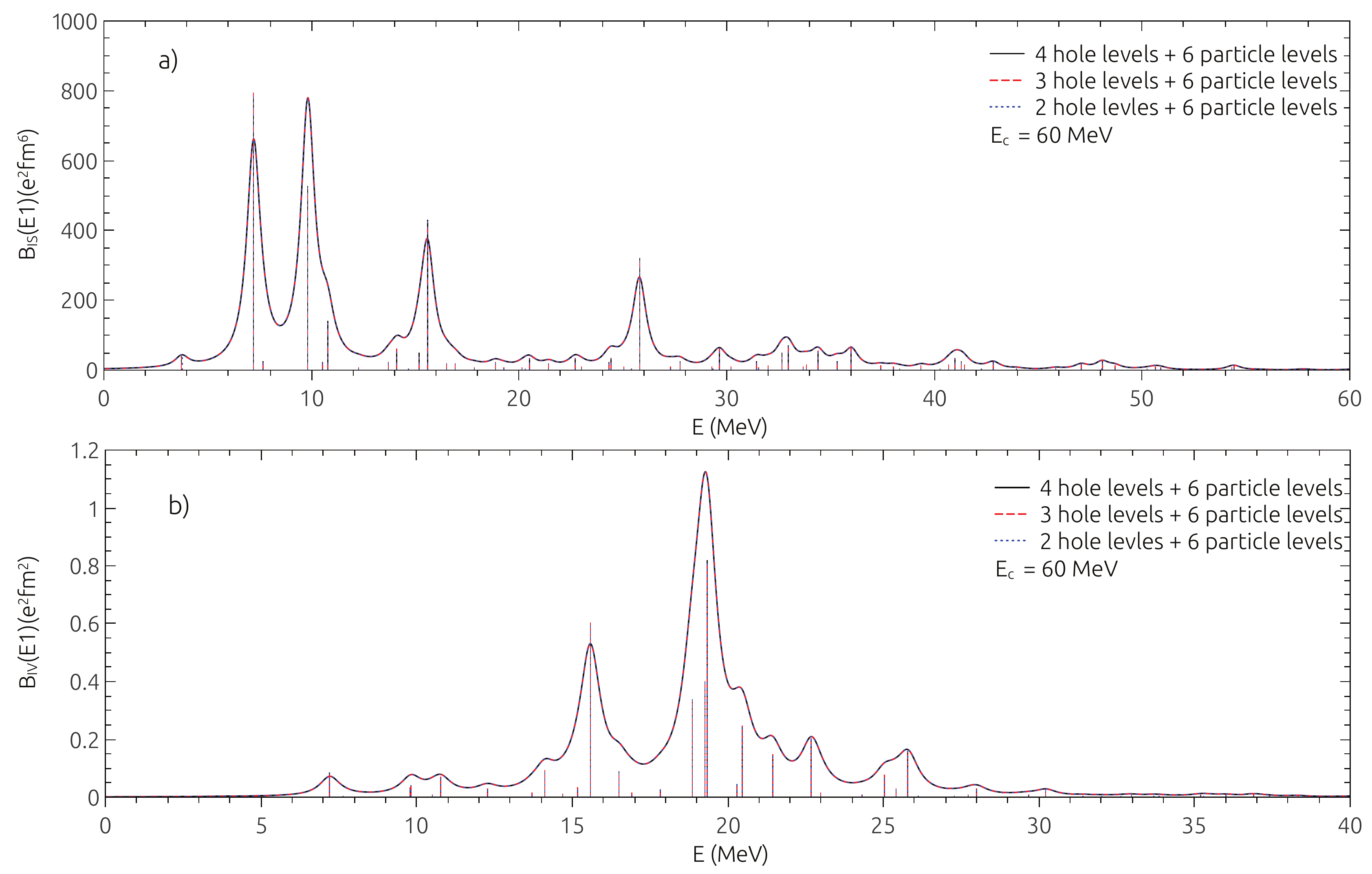}
       \caption{(Color online) The IS (a) and IV (b) dipole transition probabilities of $^{22}$O obtained within the SC-HFEPRPA by removing one and two hole levels. 
        \label{Nhole}}
    \end{figure}

\begin{figure}[h]
       \includegraphics[scale=0.45]{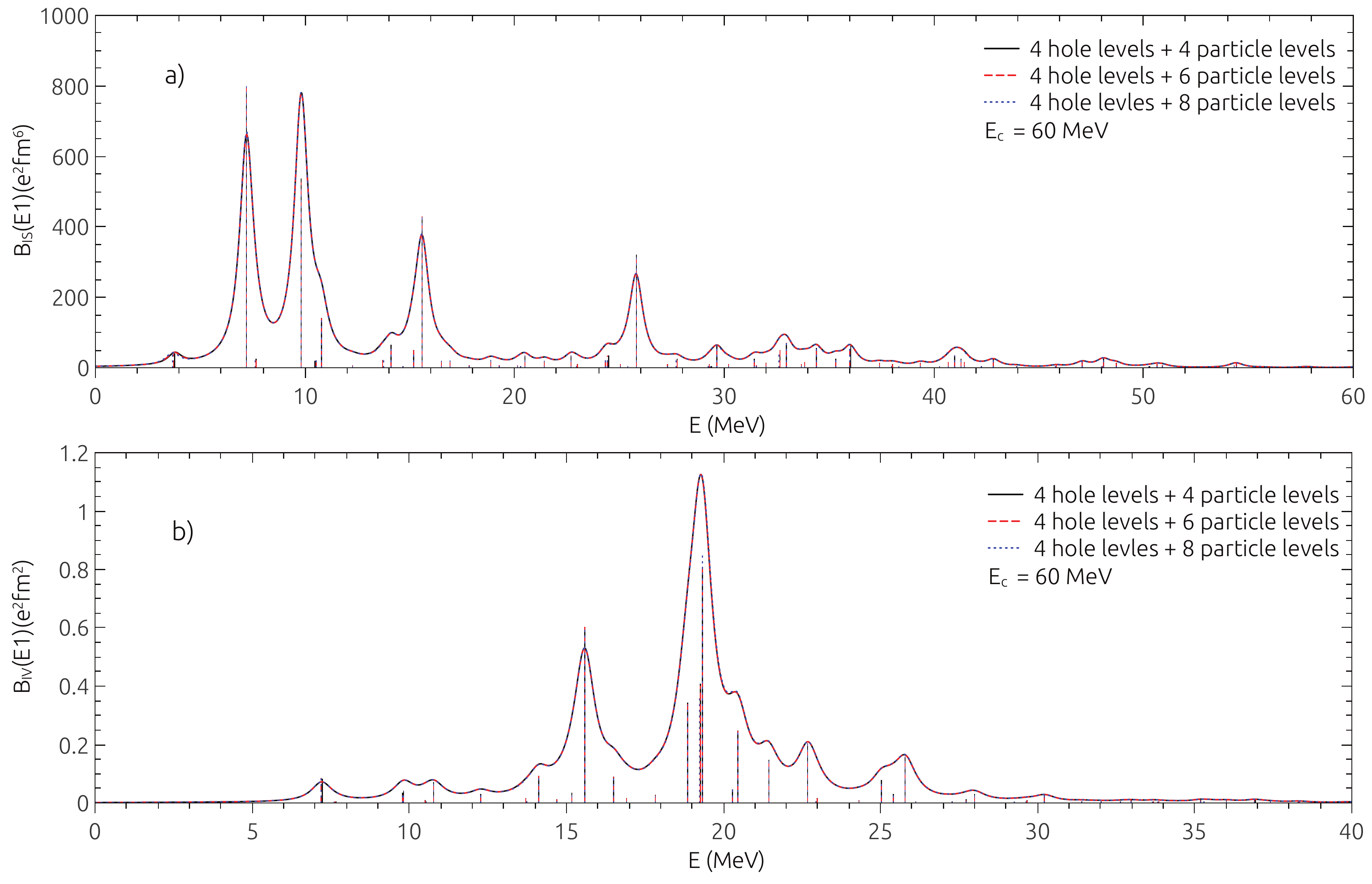}
       \caption{(Color online) The IS (a) and IV (b) dipole transition probabilities of $^{22}$O obtained within the SC-HFEPRPA by removing and adding two particle levels. 
        \label{Nparticle}}
    \end{figure}

To test the above assumption for the truncated single-particle levels, we have selected different configuration spaces by adding or removing some hole or particle levels. The numerical test for $^{22}$O with $E_c = 60$ MeV shows that removing one and two hole levels (Fig. \ref{Nhole}) or removing/adding two particle levels (Fig. \ref{Nparticle}) slightly changes the energy of the spurious state (Table \ref{table2}) but the corresponding IV and IS B(E1) distributions remain practically unchanged with different truncations. Therefore, to keep a reasonable calculation time, the truncated space for the EP calculations given in Table \ref{table1} is chosen.

\begin{figure}[h]
       \includegraphics[scale=0.45]{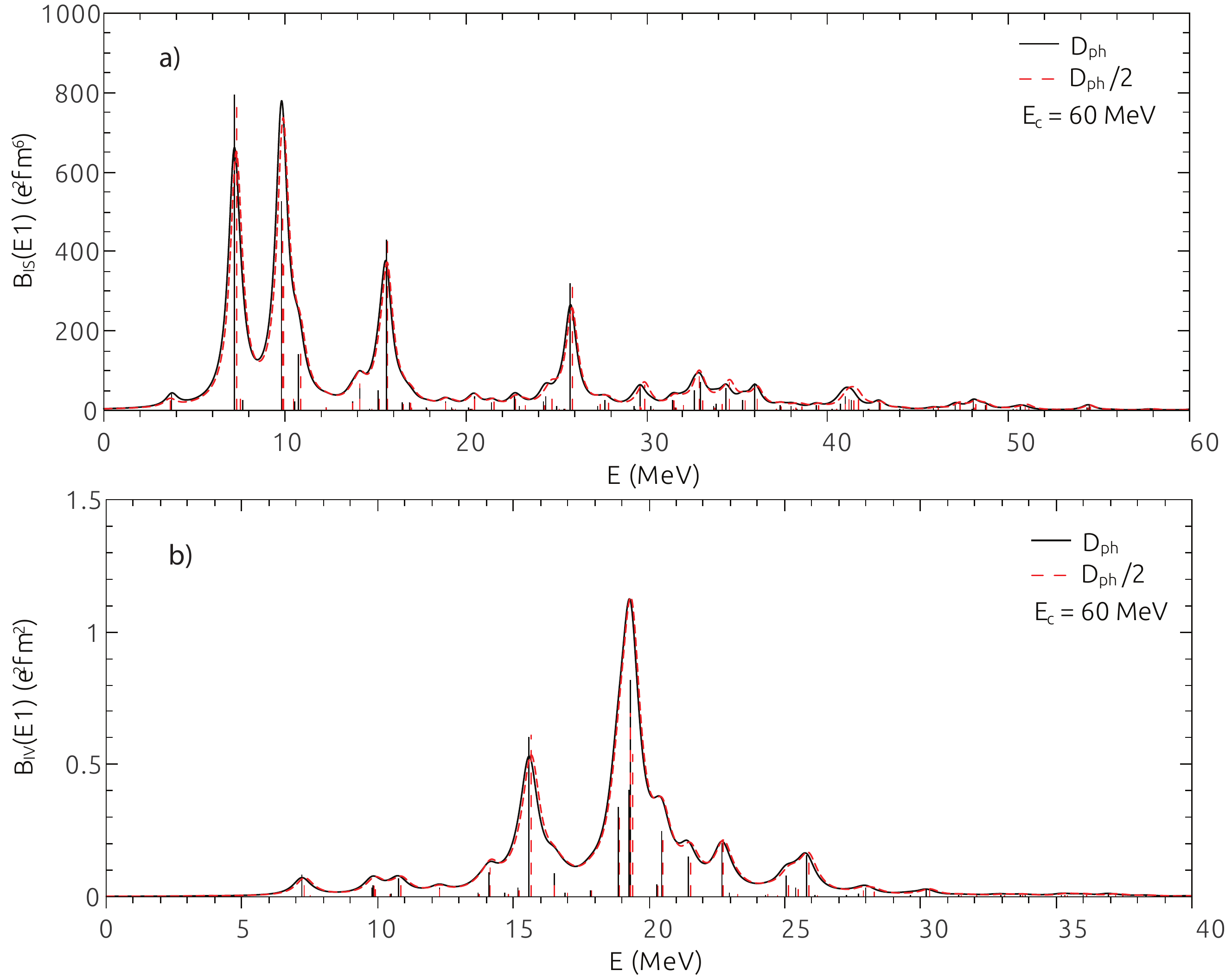}
       \caption{(Color online) The IS (a) and IV (b) dipole transition probabilities of $^{22}$O obtained within the SC-HFEPRPA by using different GSC factors $D_{ph}$. 
        \label{Dtest}}
    \end{figure}

The spurious state's energy might be also affected by the factor $\dfrac{1}{2}$, which is added to the GSC factor $D_{ph}$ as mentioned in Refs. \citep{Rowe,Len}. The numerical test for $^{22}$O with $E_c = 60$ MeV indicates that no significant change is seen between the results obtained by using $D_{ph}$ and $\dfrac{1}{2}D_{ph}$ (Fig. \ref{Dtest}). Therefore, to see the effect of GSC beyond the RPA at its strongest, we use the formalism without the factor $\dfrac{1}{2}$ in front $D_{ph}$ as has been done previously in Ref. \cite{hung2016}. 

\begin{table}[h]
\caption{Results obtained within the SC-HFEPRPA for $^{22}$O by using $E_c =$ 60 MeV and different truncated single-particle levels.}
\begin{tabular}{|c|c|c|c|c|c|c|}
\hline\hline 
Hole levels & Particle levels & Pairing strength & GSC factor & Spurious energy & IS EWSR & IV EWSR \\ 
& & (MeV) & & (MeV) & (\%) & (\%) \\
\hline 
4 & 6 & 0.376 & $D_{ph}/2$ & 3.66 & 100.54 & 100.31 \\ 
\hline 
4 & 6 & 0.376 & $D_{ph}$ & 3.75 & 100.32 & 100.07 \\ 
\hline 
3 & 6 & 0.379 & $D_{ph}$ & 3.76 & 100.32 & 100.07\\ 
\hline 
2 & 6 & 0.391 & $D_{ph}$ & 3.76 & 100.37 & 100.07 \\ 
\hline 
4 & 4 & 0.425 & $D_{ph}$ & 3.79 & 100.38 & 100.05\\ 
\hline 
4 & 8 & 0.320 & $D_{ph}$ & 3.68 & 100.20 & 100.31 \\ 
\hline\hline
\end{tabular} 
\label{table2}
\end{table}
    \begin{figure}[h]
       \includegraphics[scale=0.45]{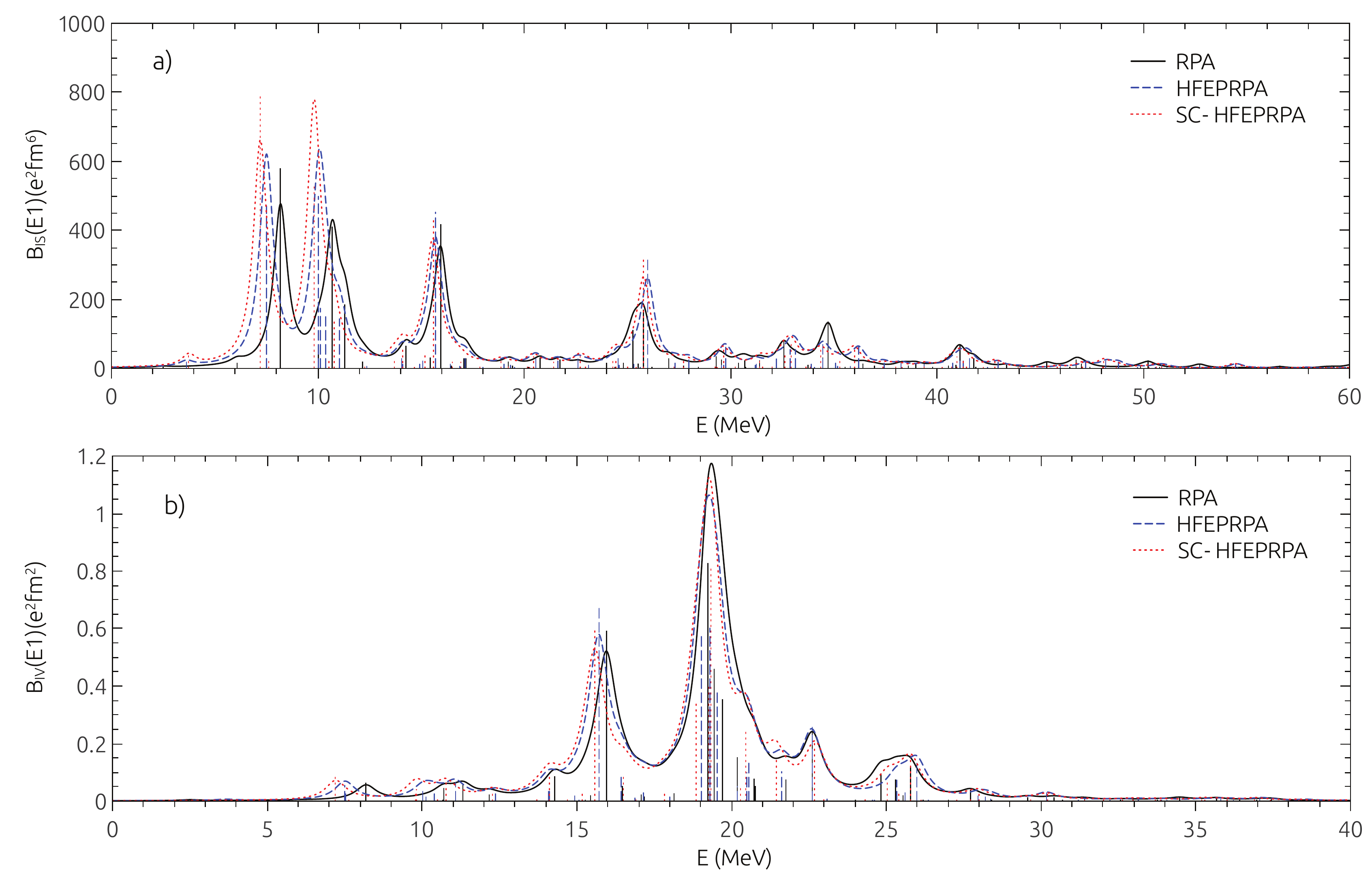}
       \caption{(Color online) The IS (a) and IV (b) dipole transition probabilities of $ ^{22} $O obtained within different approaches. The sticks and lines represent to the probabilities $B(E1,0\rightarrow 1^-)$ and strength functions $S(E)$ (the smoothing parameter $\varepsilon=0.4$ MeV), respectively. 
        \label{22O}}
    \end{figure}
    \begin{figure}[h]
       \includegraphics[scale=0.45]{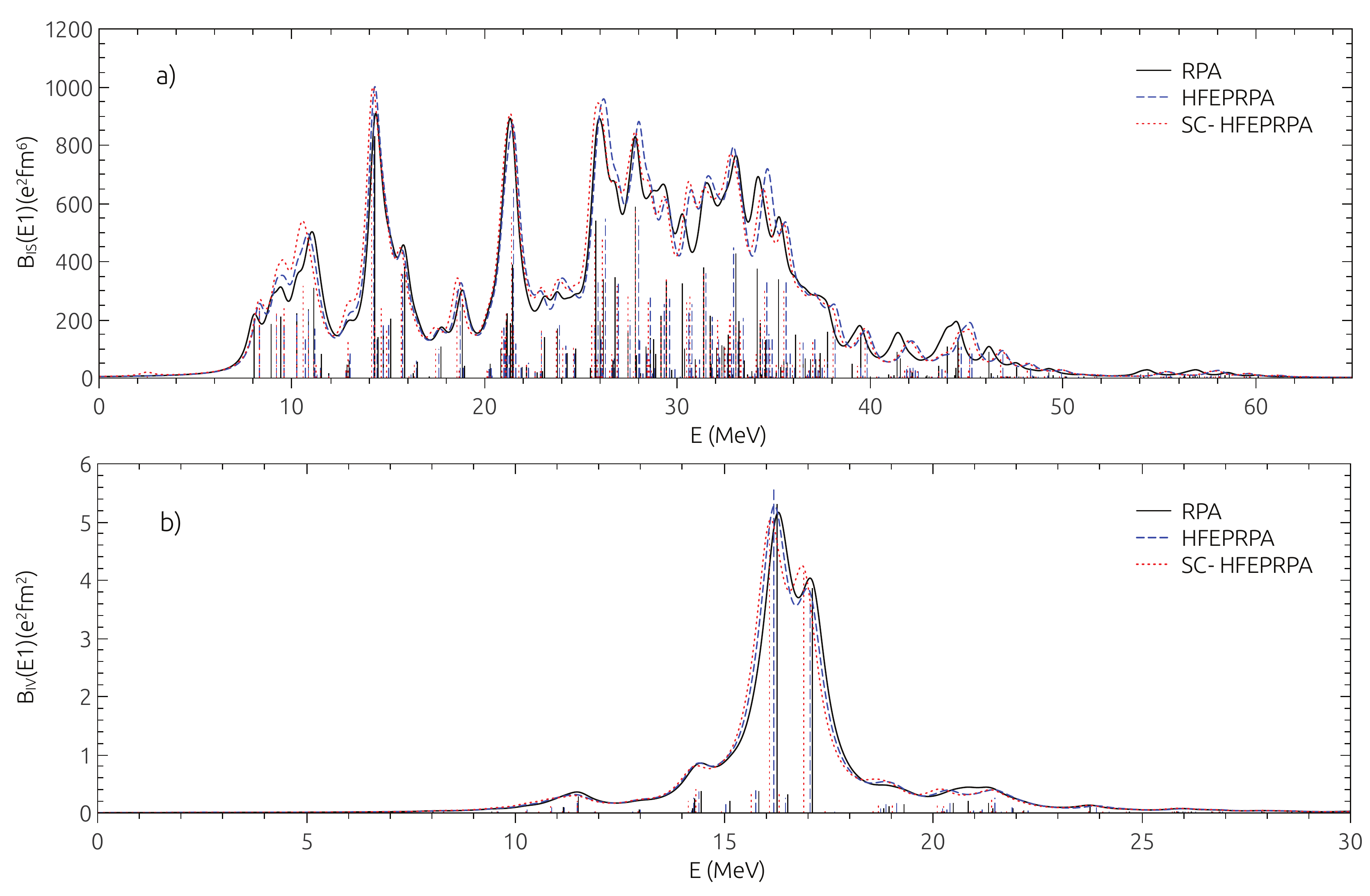}
       \caption{(Color online) Same as Fig. \ref{22O} but for $^{60}$Ni. 
        \label{60Ni}}
    \end{figure}
    \begin{figure}[h]
       \includegraphics[scale=0.45]{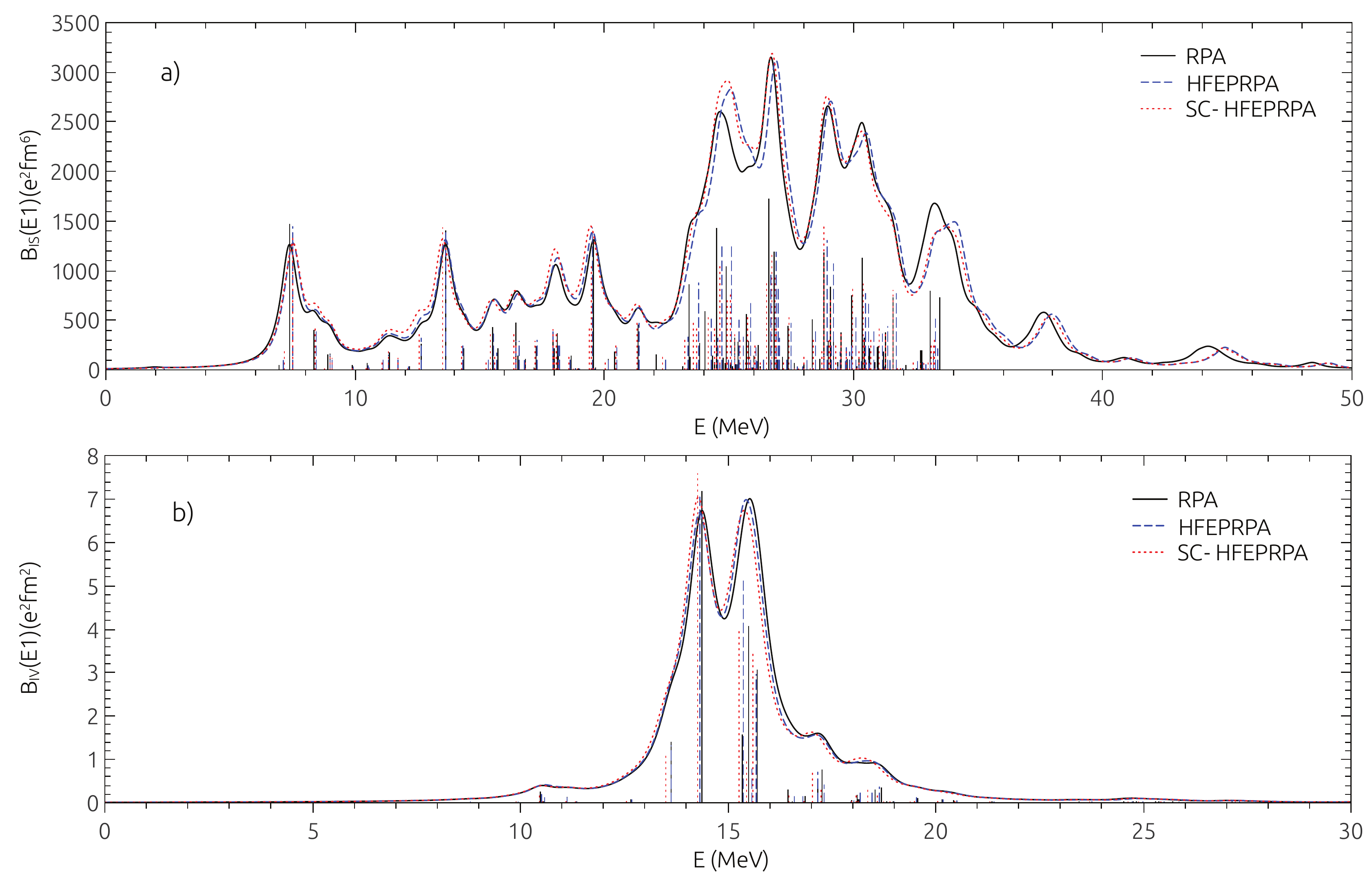}
       \caption{(Color online) Same as Fig. \ref{22O} but for $^{90}$Zr. 
        \label{90Zr}}
    \end{figure}

Shown in Figs. \ref{22O}--\ref{90Zr} are the IS and IV transition probabilities $B(E1, E_i)$ and strength functions $S(E1)$ of all nuclei under consideration obtained within the RPA, HFEPRPA, and SC-HFEPRPA for the dipole $ J^{\pi}=1^{-} $. It is seen from these figures that the strengths of the IS and IV GDRs are shifted down, similar to those reported in our previous work within the RRPA \cite{hung2016}. The shift of the strength is significant in light $ ^{22} $O nucleus (Fig. \ref{22O}), small in medium $ ^{60} $Ni nucleus (Fig. \ref{60Ni}), and insignificant in heavy $ ^{90} $Zr one (Fig. \ref{90Zr}). The explanation of this effect comes from the mean-field description and collectivity, which are good for medium and heavy nuclei because of their statistical properties, and become worse in light systems. The results obtained also show that the SC-HFEPRPA produces a stronger shift than that takes place within the HFEPRPA, whose mean field is not affected by the RPA occupation numbers. On the other hand, the spurious mode is shifted up in all cases within the HFEPRPA and SC-HFEPRPA. However, the shift is not too strong, hence the spurious mode is still well separated from the other physical states. This spurious mode is also approximately suppressed by a modified isoscalar dipole operator as in Eq. (32) of Ref. \cite{Colo2013}. In general, for medium and heavy nuclei, the results obtained by renormalizing the RPA as done in the present paper  are not much different from the predictions of the conventional RPA.

Shown in Table \ref{table3} are the fulfillments of the EWSR within the RPA, HFEPRPA, and SC-HFEPRPA. For the HFEPRPA and SC-HFEPRPA, the IS and IV EWSRs are fulfilled owing to the contribution of pairing correlation and GSC. These results confirm the assumption that the effect caused by the $pp$ and $hh$ configurations or high-order ones is effectively included in the pairing correlation. Although the EWSR is restored in both HFEPRPA and SC-HFEPRPA, the basic difference between these two approaches is that the modification of the mean field via RPA occupation numbers is performed only within the SC-HFEPRPA. On the other hand, unlike some common implementations of QRPA, the particle number is always conserved in the present EP-based approaches.

\begin{table}[h]
\caption{The fulfillment of the IS and IV EWSRs obtained within the RPA, HFEPRPA, and SC-HFEPRPA by using the MSk3 interaction with the cut-off energy $ E_c=60 $ MeV.}
\begin{tabular}{|c|c|c|c|c|c|c|c|c|c|c|c|c|}
\hline\hline
&\multicolumn{3}{c|}{IS$(\%)$} & \multicolumn{3}{c|}{IV$(\%)$} \\
\hline
	&RPA&HFEPRPA&SC-HFEPRPA&RPA&HFEPRPA&SC-HFEPRPA\\
\hline
    $^{22}$O& 99.611 & 99.556 & 100.325 & 99.920 & 100.648 & 100.074 \\
    $^{60}$Ni& 99.659 & 100.991 & 101.161 & 99.962 & 99.087 & 99.229 \\
   $^{90}$Zr& 99.565 & 100.312 & 100.791 & 99.964 & 98.997 & 99.081 \\
    \hline\hline
\end{tabular}
\label{table3}
\end{table}

During the renormalization, the pairing effect is found to be reduced in the SC-HFEPRPA. As presented in Table \ref{table4}, the reduction of pairing energy, which is always more than 10$\%$ and reaches $\sim 30\%$ in the neutron-rich $ ^{22} $O nucleus. This reveals the mutual influence of pairing and residual correlations via the GSC factor $ D_{ph}^{EP} $ and RPA occupation numbers. Both of the short-range pairing and long-range \textit{ph} residual interactions are included in the nuclear mean field, which may serve as an explanation for this phenomena. In the beginning, the mean field (HFEPRPA) contains only the pairing correlation. After the SC-HFEPRPA calculation was performed, the new mean field contains both of the pairing and residual correlations. Therefore, the pairing correlation is reduced to give room for the residual correlation, which comes from the RPA. In particular, the mean field is modified not only by pairing, as within the HFEPRPA, but also by the residual correlation from the RPA within the SC-HFEPRPA in a self-consistent way. This reduction of pairing can be associated with the anti-pairing effect within the SC-HFEPRPA.

\begin{table}[h]
\caption{Pairing energies obtained within the HFEPRPA and SC-HFEPRPA. The quantity  $ \delta=(E_{pair}^{HFEPRPA}-E_{pair}^{SC-HFEPRPA})/E_{pair}^{HFEPRPA} $ (\%) represents the depletion of pairing effect.}
\begin{tabular}{|c|c|c|c|c|}
\hline\hline
&\multicolumn{3}{c|}{E$ _{pair} $ (MeV)} \\
\hline
	&HFEPRPA&SC-HFEPRPA& $\delta$(\%)\\
\hline
    $^{22}$O & -4.623 & -3.312 & 28.36\%\\
    $^{60}$Ni &  -3.301 & -2.937 & 11.02\%\\
   $^{90}$Zr & -1.290 & -1.095 & 15.12\%\\
    \hline\hline
\end{tabular}
\label{table4}
\end{table}
\begin{table}[h]
\caption{Ratio $r = S_{PDR}/S_{GDR}$ obtained within the RPA, HFEPRPA, and SC-HFEPRPA.}
\begin{tabular}{|c|c|c|c|}
\hline\hline
&~RPA~&~HFEPRPA~&~SC-HFEPRPA~ \\
\hline
	$^{22}$O&0.0236&0.0297&0.0301 \\
	$^{24}$O&0.0565&0.0627&0.0623 \\
	$^{28}$O&0.1030&0.1085&0.1110 \\
	$^{60}$Ni&0.0272&0.0263&0.0269 \\
	$^{90}$Zr&0.0156&0.016&0.0155 \\
\hline\hline
\end{tabular}
\label{table5}
\end{table}
    \begin{figure}[h]
       \includegraphics[scale=0.6]{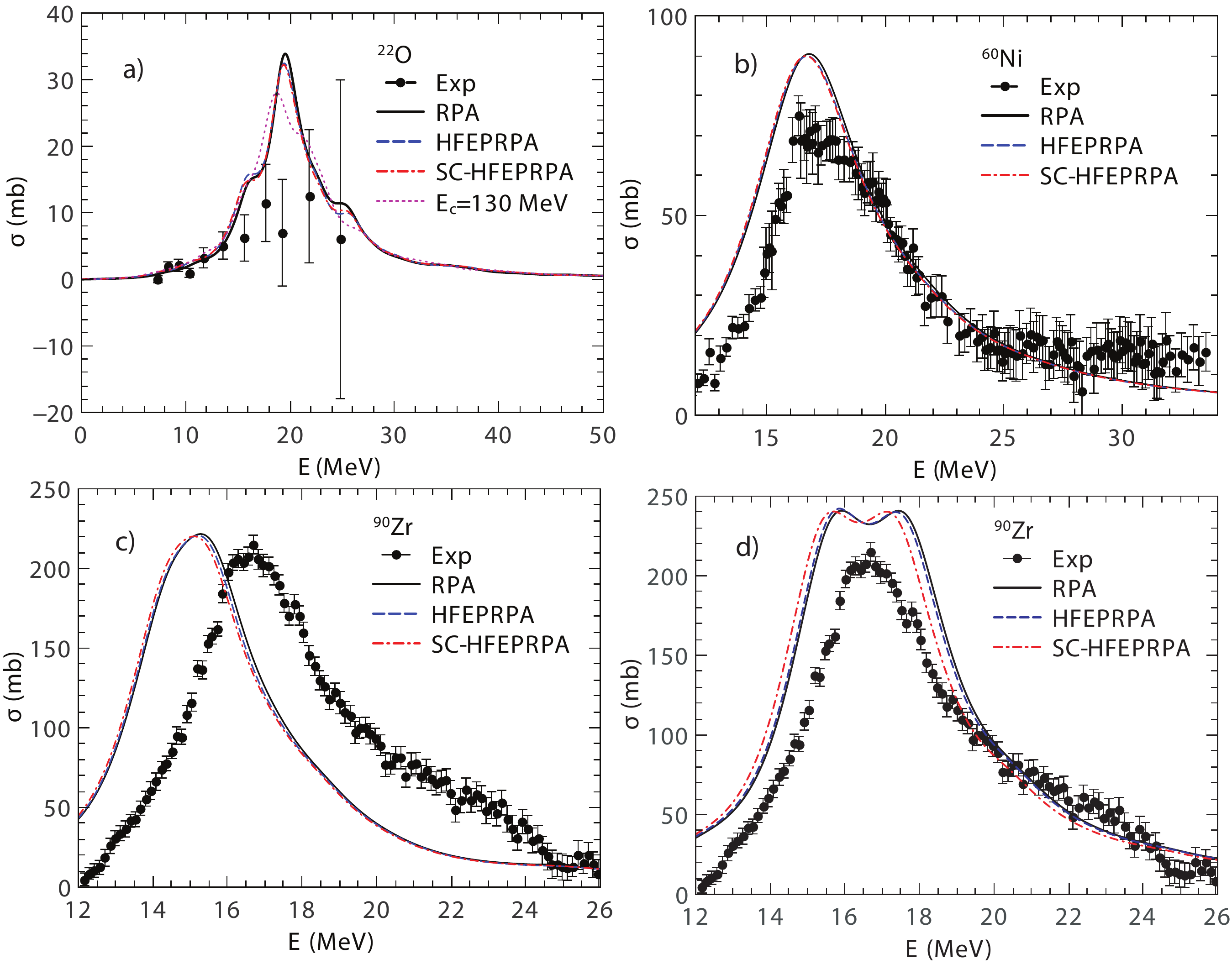}
       \caption{(Color online) GDR cross-sections for $^{22}$O, $^{60}$Ni, and $^{90}$Zr obtained within different methods. Results in (a), (b), and (c) are obtained by using the MSk3 interaction, whereas those in (d) are obtained by using the SkM$^*$ interaction. The experimental data for $^{22}$O, $ ^{60} $Ni, and $ ^{90} $Zr are taken from Refs. \cite{Lei,Lep,Ful}, respectively. The dotted line in (a) is predicted by the SC-HFEPRPA by using $E_c=$ 130 MeV. 
        \label{cs}}
    \end{figure}

The PDR, which appears around the particle-emission threshold in the neutron-rich nuclei, is also observed in our calculations for $ ^{22,24,28} $O, $ ^{60} $Ni, and $ ^{90} $Zr. Shown in Table \ref{table5} is the ratio $ r = S_{PDR}/S_{GDR}$ between the EWSS of the PDR ($ 0\leq E^{\nu}\leq 12 $ MeV) and GDR ($ 0\leq E^{\nu}\leq 60 $ MeV). The interval of PDR energies is often chosen from 0 up to 15 MeV \cite{Dang2001,Hung2013}. We choose $E_{max}=$ 12 MeV to avoid the overlap of the PDR to the GDR region, which is seen in the Figs. \ref{22O}, \ref{60Ni}, and \ref{90Zr}. The values of $ r $ are small (1--2$ \% $) in the stable nuclei $ ^{60} $Ni and $ ^{90} $Zr, whereas its values increase from $ 3\% $ to $ 11\% $ in the neutron-rich nuclei $ ^{22,24,28} $O. Both the HFEPRPA and SC-HFEPRPA produce a pronounced PDR in these neutron-rich nuclei. This enhancement can be explained by the contribution of EP, leading to the GSC factor $ D_{ph}^{EP} $, as has been discussed in Ref. \cite{Hung2013}.

Finally, we compare our calculated photoabsorption cross-sections in three nuclei under consideration with the experimental data \cite{Lei,Lep,Ful} in Fig. \ref{cs}. The calculated cross-sections are generated from the strength function $S(E1)$ obtained by using the MSk3 and SkM$^*$ interactions. The smoothing parameter $\varepsilon$ in Eq. (\ref{strength}) is chosen equal to 1.5, 2.5, and 1.25 to produce the GDR width $\Gamma=$ 3.0, 5.0, and 2.5 MeV for $^{22}$O, $^{60}$Ni, and $^{90}$Zr, respectively. The results obtained show no significant difference between the conventional RPA and our approaches. The maximum peak of cross-sections obtained within our approaches is shifted down to the lower excitation energy. The shift is the most prominent for $^{90}$Zr, which is around 2 MeV [Fig. \ref{cs}(c)].  When the SkM$^{*}$ interaction, which is known to well describe the GDR in heavy nuclei \cite{Ina}, is used for $^{90}$Zr, the shift is eliminated and a better agreement with the experimental data is seen [Fig. \ref{cs}(d)]. It can also be observed from Figs. 10 (c) and 10 (d) that, for $^{90}$Zr nucleus, the SkM$^{*}$ interaction produces an enhancement of the IV dipole transition probabilities as compared to those obtained by using the MSk3 interaction. This comparison shows that, for the giant resonances, the difference between the predictions by various approximations such as RPA, HFEPRPA, and SC-HFEPRPA is smaller than those caused by different interactions, so the agreement is fair if the interaction is well tailored to the phenomenon at hand.
\section{Conclusions} 
The present paper proposes an approach to renormalize the RPA making use of the exact pairing solution. The GSC factor, which includes pairing correlation, is employed to renormalize the residual interaction. The calculations are performed at each separate multipolarity $J^{\pi} $ in two ways, the non-self-consistent (HFEPRPA) and self-consistent (SC-HFEPRPA), for $ ^{22,24,28} $O, $ ^{60} $Ni, and $ ^{90} $Zr nuclei by using the Skyrme interaction MSk3. The results obtained show that the drawback of the $ ph $RRPA is removed, namely the IS and IV EWSRs are fulfilled without adding any $pp$ and $hh$ configurations, hence the extension of RPA matrices and time-consuming calculations are avoided. As compared to the RPA results, the effects of GSC and EP in the renormalization is significant in light nuclei and small in medium and heavy nuclei.

The anti-pairing effect is observed the first time within the SC-HFEPRPA, which reduces the pairing energy from more than $ 10\% $ up to around $ 30\% $ in the neutron-rich $ ^{22} $O nucleus. This shows the contribution from the mutual effect of the short-range pairing and long-range \textit{ph} residual interaction to the mean field. The PDR, owing to the oscillation of the excess neutron against the proton-neutron core, is also found to be enhanced in neutron-rich nuclei because of the pairing effect. The GDR cross-sections are also calculated and no significant difference between the results obtained within the conventional RPA and our approaches is seen. They agree fairly well with the experimental data.

\acknowledgments
L.T.P. acknowledges the International Internship Program at RIKEN, where a part of this work was carried out. This work is funded by the Vietnam Government under the Program of Development in Physics toward 2020 (Grant No. DTDLCN.02/19) and the National Foundation for Science and Technology Development (NAFOSTED) of Vietnam (Grant No. 103.04-2017.69).

\end{document}